\documentclass[conference]{IEEEtran}
\IEEEoverridecommandlockouts
\usepackage{stfloats}
\usepackage{cite}
\usepackage{amsmath,amssymb,amsfonts}
\usepackage{algorithmic}
\usepackage{graphicx}
\usepackage{textcomp}
\usepackage{xcolor}

\def\BibTeX{{\rm B\kern-.05em{\sc i\kern-.025em b}\kern-.08em
    T\kern-.1667em\lower.7ex\hbox{E}\kern-.125emX}}
\begin{document}

\title{TrafPS: A Visual Analysis System Interpreting Traffic Prediction in Shapley\\
}

\author{\IEEEauthorblockN{Yifan Jiang\IEEEauthorrefmark{1},
		Zezheng Feng\IEEEauthorrefmark{1}\IEEEauthorrefmark{2},
		Hongjun Wang\IEEEauthorrefmark{1}, 
		Zipei Fan\IEEEauthorrefmark{3},
		Xuan Song\IEEEauthorrefmark{1}}
	\IEEEauthorblockA{\textit{\IEEEauthorrefmark{1}Department of Computer Science and Engineering,Southern University of Science and Technology} \\
		Email: \IEEEauthorrefmark{1}yifanjiang921@gmail.com,\IEEEauthorrefmark{1}\{11851004@mail., wanghj2020@mail., songx\}sustech.edu.cn\\
		\IEEEauthorrefmark{2}The Hong Kong University of Science and Technology \\
		Email: \IEEEauthorrefmark{2}zfengak@connect.ust.hk\\
		\IEEEauthorrefmark{3}University of Tokyo \\ 
		Email: \IEEEauthorrefmark{3}fanzipei@iis.u-tokyo.ac.jp\\
}}

\maketitle


\begin{abstract}
In recent years, deep learning approaches have been proved good performance in traffic flow prediction, many complex models have been proposed to make traffic flow prediction more accurate. However, lacking transparency limits the domain experts on understanding when and where the input data mainly impact the results. Most urban experts and planners can only adjust traffic based on their own experience and can not react effectively toward the potential traffic jam. To tackle this problem, we adapt Shapley value and present a visualization analysis system , which can provide experts with the interpretation of traffic flow prediction. TrafPS consists of three layers, from data process to results computation and visualization. We design three visualization views in TrafPS to support the prediction analysis process. One demonstration shows that the TrafPS supports an effective analytical pipeline on interpreting the prediction flow to users and provides an intuitive visualization for decision making. 
\end{abstract}


\section{Introduction}

As increasingly boost the volume of vehicles increases and the travel demand of citizens, the traffic in the urban area has becoming more and more complex. Traffic flow prediction has been an important problem to urban planners, which can help them estimate the changes of the traffic so that they can respond in time to possible traffic jams or congestion~\cite{b2}. To get a more accurate prediction, many approaches have been explored~\cite{b1,b3,b4}. With its strong forecasting performance, deep learning has proved sound in traffic flow prediction and various complex models have been proposed to make traffic flow prediction more accurate.


Even though selecting proper deep learning models may bring high accuracy to the prediction, these complex models are always regarded as black boxes that lack transparency, which leads to the experts having no idea how the models impact the prediction. Without any interpretations of the model, domain experts cannot react effectively to predictions such as when and where the input data mainly impact the results.

To solve this problem, inspired by other urban analysis studies~\cite{b7,b8,b9,b10}, we want to build a visual analysis system to provide prediction with interpretation and additional related information. As the system needs an interpretation method, we employ the Shapley value, proposed by Shapley~\cite{b5} and adopted in~\cite{b6}.  The Shapley value is widely considered a unique unbiased approach to fairly allocating the total award a coalition to each input feature.  In the traffic flow prediction task, we use the Shapely value to quantify the correlations between the prediction results in the selected region and actual traffic flow or geography area. Then we propose TrafPS, a visual analysis system interpreting traffic prediction in Shapley, to enhance the urban expert on analyzing traffic prediction and reacting to potential traffic jams.

TrafPS contains three layers with different functions to provide efficient analysis: i) Data processing layer preprocess raw data and prepare for the other layers (e.g., data partition and make division). ii) Prediction-Interpretation layer consists of two computation models. Prediction Model computes the traffic flow prediction in future timestamps, and Interpretation Model calculates the corresponding interpretation. iii)Visualization analytical layer visualizes results in former layers in several interactive views to provide the visualization-assisted analysis (e.g., Map-Trajectory View, Radar Glyph View, and Fine-grained Grid View).

We implement TrafPS on a real-world taxi data set in Chengdu and will present the usage of TrafPS through demonstration. Our demonstration will invite ICDE attendees to act as urban analysts to analyze current traffic flow and future predictions. Attendees can operate on the Visualization analytical layer interface to detect traffic jams and do appropriate reactions to prevent them based on the interpretation provided by Prediction-Interpretation layer.


The contributions of this work are as follows:
\begin{itemize}
\item We introduce Shapley value to the traffic flow prediction task to interpret the prediction made by complex machine-learning models. 

\item We proposed an interactive visual analytics system named \emph{TrafPS} provides six visual components to support analysts in understanding the prediction and make effective reaction. 

\item We designed a radar glyph to simply and informatively summarize prediction results and their interpretation in the Shapely value for the selected area.
\end{itemize}

We demonstrate the framework of TrafPS in Section \uppercase\expandafter{\romannumeral2} and introduce three visualization views in Section \uppercase\expandafter{\romannumeral3}. Then we show TrafPS's usage and how it works in Section \uppercase\expandafter{\romannumeral4} and conclude in Section \uppercase\expandafter{\romannumeral5}.

\section{Model}

\begin{figure}[h]
	\centering
	\includegraphics[width=0.48\textwidth]{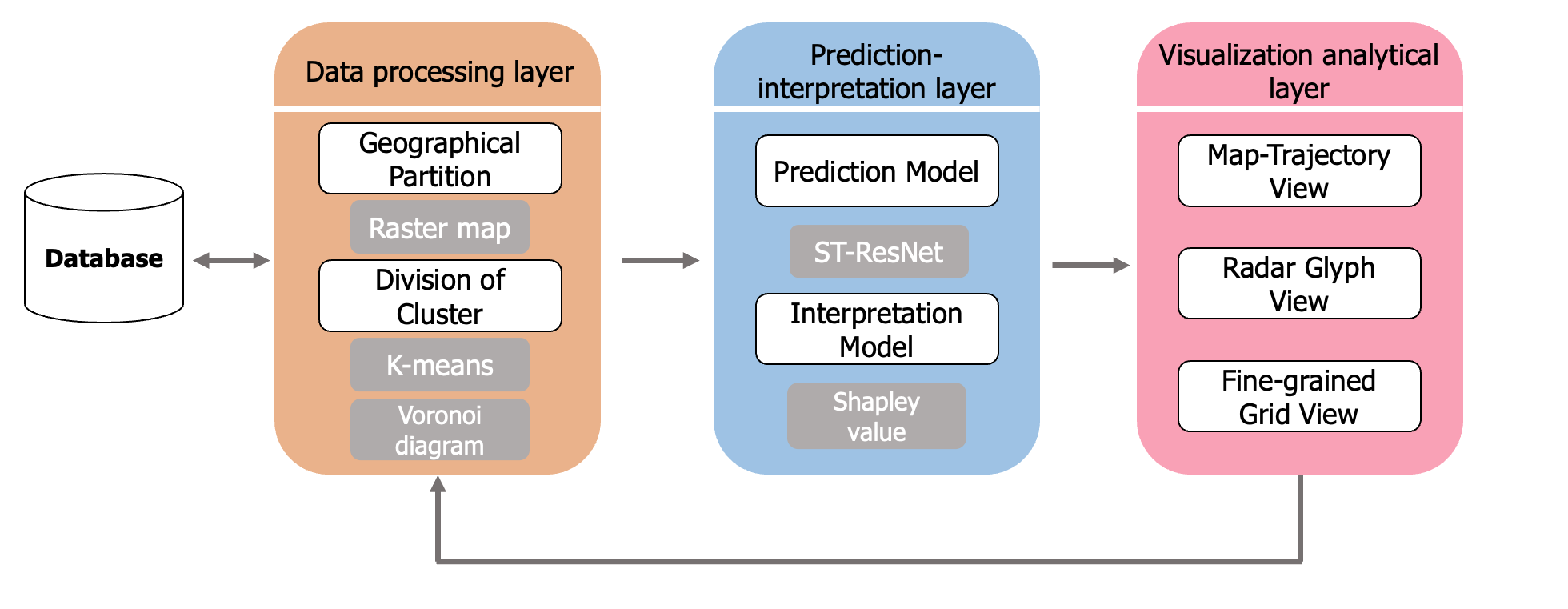}
	\caption{An overview of the TrafPS system, which mainly consists of three layers: Data processing layer, Prediction-interpretation layer and visualization analytical layer}
	\label{fig:system}
	\vspace{-1em}
\end{figure}

In this section, we describe the TrafPS system framework in \fzz{Fig.~\ref{fig:system}}. It consists of three layers:  (i) Data processing layer; (ii) Prediction-Interpretation layer; and (iii) Visualization analytical layer.  We build the front-end of TrafPS on Vue\footnote{https://vuejs.org/}. and the back-end on Flask\footnote{https://flask.palletsprojects.com}. 

The input dataset of this work consists of the following data in Chengdu, China. \noindent\textbf{Traffic trajectory data} contains the trajectory data in Chengdu city~(Data comes from Didi Chuxing GAIA Initiative\footnote{https://gaia.didichuxing.com}.) from 10/01/2016 to 11/31/2016. \noindent\textbf{Road network data} contains the traffic intersection data in the same location of Chengdu city~(Data comes from OSMnx\footnote{https://www.openstreetmap.org/}.).

\subsection{Data processing layer}

Data processing layer is the basic layer of TrafPS. It takes real-world datasets and does necessary preparation for the Prediction-Interpretation layer (e.g., data cleaning, data division, and data aggregation).

\subsubsection{\textbf{Geographical Partition}}
When adopting ST-ResNet~\cite{b2} as the prediction method, the input of the model should be a sequence of fixed-sized matrices. Therefore, we should process the data to fit the input of it. We first rasterize the city into grids to provide enough spatial and temporal information of input data. Then we count the passing trajectories for each grid to get the in-/out-flows.

\subsubsection{\textbf{Division of Cluster}}
After partitioning the city into grids as the single unit, we then cluster them based on the number of roads in each grid and match each grid with the clusters it belongs to which reduce the imbalance caused by the great difference of the number roads in different grids. Therefore, we run kmeans algorithm to divided the intersections in groups. We eventually choose 21 as the number of the cluster as we want more number of cluster to provide more local and detailed information after trying 0 to 27.
Moreover, we employ Voronoi diagram to generate each intersection with a polygon which aims to get the border of each cluster so that each grid can match their own clusters and let visualization system exhibit the location of each cluster.

\subsection{Prediction-interpretation layer}

Prediction-Interpretation layer is the core layer of TrafPS that runs Prediction Model and Interpretation Model to support the system's analysis function. Interpretation Model takes the result of  Prediction Model and computes its corresponding interpretation.

\subsubsection{\textbf{Prediction Model}}~\label{prediction}
In this task, Prediction  Model use ST-ResNet~\cite{b1} as our machine-learning model to get the prediction. Data processing layer gets our flow matrix by dividing data every 10 minutes, and the prediction task uses data in 5 continuous  timestamps to predict the timestamp next.

\subsubsection{\textbf{Interpretation Model}}~\label{interpretation}
Based on the Additive Feature Attribution Methods~\cite{b6}, Interpretation Model assumes the flow prediction of one cluster as the summation of other areas' contributions. Each time interpreting one cluster's prediction, Interpretation Model computes the correlation between the cluster's neighbor area and the prediction result. The larger the area's correlation value, the more positive the area contributes to the prediction result. For instance, an area with a large correlation pushes the model to give a high value during the prediction of the cluster being interpreted. 

Traffic flow in grids tends to fluctuate, and sometimes several trajectories can considerably impact its prediction. Thus, when interpreting the flow prediction of one grid, Interpretation Model will compute the correlation between the existing trajectories and the grid's prediction. 

\begin{figure*}[h]
	\centering
	\includegraphics[width=1\textwidth]{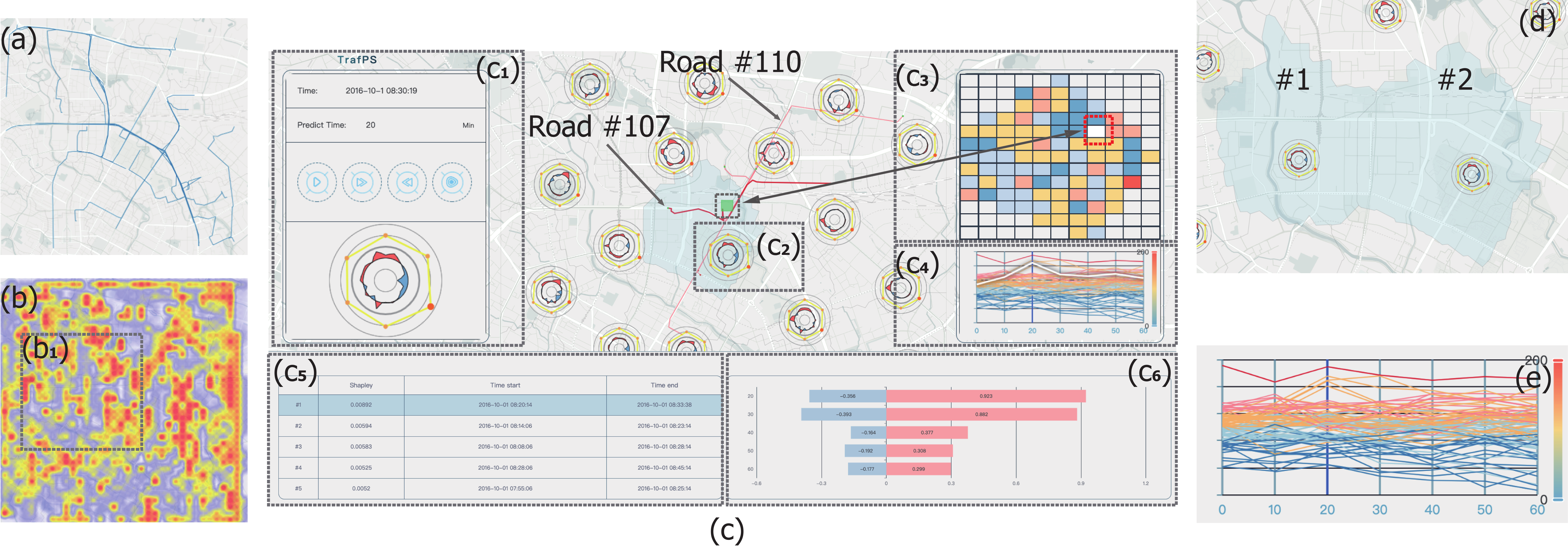}
	\caption{The visualization view (c) of TrafPS consists of: A dashboard ($c_1$) on the left of TrafPS, which can select trajectory data (a) in a specific timestamp and project heatmap (b); Radar glyphs ($c_2$) on each cluster, and a magnified one on the left ($c_1$); A Square Chart ($c_3$) and a Parallel Coordinates Plot ($c_4$) visualize the flow data in grids. TrafPS will demonstrate the five most significant trajectories on the map with a table list ($c_5$) and a  Bi-directional Column Chart ($c_6$) at the bottom after selecting a grid.}
	\label{fig:case}
\end{figure*}

\subsection{Visualization analytical layer}

Visualization analytical layer is the user interface of the TrafPS. It visualizes data or model results from the former two layers with informative and straightforward visualization views. Each view supports several interactive operations, e.g., zoom to scale the size, keyboard input, or mouse select. Users can switch between views and locate a specific target at ease. We will introduce these views in this layer in the next section.

\section{Visualization Design}

\begin{figure}[h]
	\centering
	\includegraphics[width=0.5\textwidth]{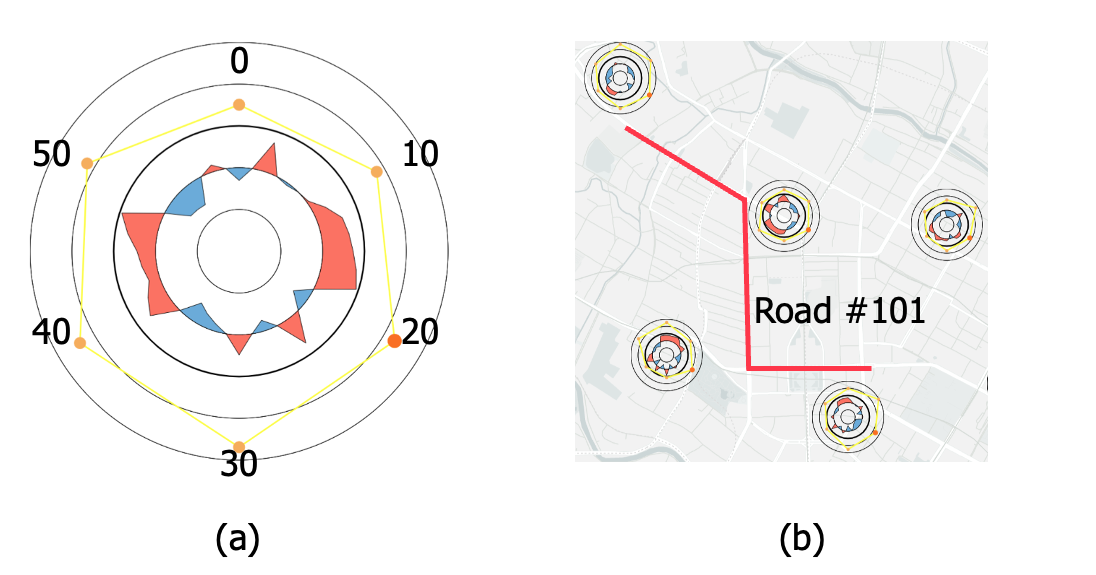}
	\caption{The design of (a) Radar Glyph View and (b) an example of how to find key roads}
	\label{fig:glyph}
\end{figure}

This section introduces three interactive and interpretation-assisted visualization views in TrafPS that ease and improve experts' prediction analysis process. For each View, we explain the purpose of visualization design and present the visualization result.

\subsection{Map-Trajectory View}

At the beginning of the analysis, having a global view of the current traffic flow can be vital as experts can be aware of the moving trend of overall traffic. An overview of traffic flow prediction is also helpful to show experts where traffic jams may happen. Thus, Map-Trajectory View visualizes the vehicle trajectory data on the geography map dynamically (\fzz{Fig.~\ref{fig:case}($a$)}). We adapt heatmap to distinguish prediction from the current traffic flow and presents data in high comparability (\fzz{Fig.~\ref{fig:case}($b$)}), where darker color means more heavy traffic flow.  Compared with other existing trajectory systems, our Map-Trajectory View provides several operators in a dashboard  (\fzz{Fig.~\ref{fig:case}($c_1$)}) to support the analysis process. Experts can pause at any timestamp to analyze traffic flow in detail or switch to locate specific timestamps. 

\subsection{Radar Glyph View}

In addition to gaining an insight into the entire traffic flow and prediction, experts also need the interpretation of prediction to understand and react effectively. Radar Glyph View places a newly designed radar glyph at the centroid of each cluster area to summarize the traffic flow prediction of the whole cluster and its corresponding interpretation (\fzz{Fig.~\ref{fig:case}($c_2$)}).  Clicking the cluster can magnify the glyphs and gain a clearer view (\fzz{Fig.~\ref{fig:case}($c_1$)}). 

\noindent\textbf{Traffic Flow Prediction:} A line chart around the glyph visualizes the traffic flow prediction in the future 50 minutes with every 10 minutes a dot point clockwise (\fzz{Fig.~\ref{fig:glyph}($a$)}). Points far away from the glyph's center represent a large prediction value, and the radar glyph highlights the point of the timestamp being interpreted. For example, the line chart in \fzz{Fig.~\ref{fig:glyph}($a$)} shows that the traffic flow will increase in the following 30 minutes, and the timestamps in 20 minutes later will be interpreted.

\noindent\textbf{Corresponding Interpretation:}  In the inner part, two diverging polygon distributions visualize the correlation between the cluster's neighbor area and the cluster's traffic flow prediction by the geographical directions of these surrounding areas—the bigger the polygon, the more powerful the positive (\textcolor{red}{red}) or negative (\textcolor{blue}{blue}) correlation. For instance, \fzz{Fig.~\ref{fig:glyph}($a$)} shows that the traffic flow will increase 20 minutes later, and the surrounding area at the west and the east may contribute to this increase. Experts can restraint the traffic flow or adjust the traffic network in those areas tentatively to avoid traffic jams. Analyzing radar glyphs in adjacent clusters can find some significant areas related to these clusters' traffic flow. For instance, the interpretation part of three radar glyphs in \fzz{Fig.~\ref{fig:glyph}($b$)} indicates Road \#101 may play an essential role in impacting local traffic flow.

\subsection{Fine-grained Grid View}

Radar Glyph View gives interpretation in a general view based on surrounding areas. Sometimes experts need to analyze the traffic flow prediction in specific places and make reactions, e.g., the main intersection in urban. In this view, we divide each cluster into grids and provide fine-grained grid-based traffic flow analysis through several visualization components.

\noindent\textbf{Square Chart~(\fzz{Fig.~\ref{fig:case}($c_3$)})}~visualizes grids in different colors based on their current traffic value. Red color means large flow value, and blue means small. Square Chart can reflect each grid's spatial information and the diversity of current traffic flow.

\noindent\textbf{Parallel Coordinates Plot~(\fzz{Fig.~\ref{fig:case}($c_4$)})}~draws each grid's traffic flow change in future 60 minutes with the same color in Square Chart (\fzz{Fig.~\ref{fig:case}($c_3$)}). The plot can accurately reflect traffic flow trends, and clicking the grid in Square Chart will highlight the corresponding time series. 


Regarding reducing traffic flow in some specific places, adjusting the traffic flow on several roads can be far more efficient than on an entire area.  In this view, experts can select a grid in Square Chart to analyze~(highlighted in \fzz{Fig.~\ref{fig:case}($c_3$)}), and the top five trajectories that have the most correlation or impact on the traffic flow prediction of this grid will appear on the map. Trajectory's color means the nature of correlation, as same as the color in radar glyph.  A list containing each trajectory's detailed data will appear (\fzz{Fig.~\ref{fig:case}($c_5$)}), as long as a Bi-directional Column Chart aggregates the correlation of each trajectory into time channels~(\fzz{Fig.~\ref{fig:case}($c_6$)}).

\section{The Demonstration}

This demonstration aims to provide an interpretation view to understand models' predictions and extract the latent patterns from traffic data. User can act as an urban analyst to analyze current traffic flow and future prediction then react to potential traffic jams with the help of interpretation provided by the demo. We perform the demo on a real-world traffic dataset that contains one-week taxi trajectory data in Chengdu, China. The demo presents the corresponding interpretation to the traffic flow prediction of 20 minutes later.

The user first operates the dashboard~(\fzz{Fig.~\ref{fig:case}($c_1$)}) to select which time period to analyze. Assume the user selects an early morning on a hoilday. Map-Trajectory View will generate moving trajectories of all real-time data (\fzz{Fig.~\ref{fig:case}($a$)}). Trajectories present a tendency to move towards the center of the map, one of the main roads connecting the urban area and the suburban area, indicating traffic jams may happen. Then he projects a heatmap~(\fzz{Fig.~\ref{fig:case}($b$)}) to see traffic flow prediction 20 minutes later, which reveals several areas with the darkest color will have congestions, including the center of the map~(\fzz{Fig.~\ref{fig:case}($b_1$)}).

Being aware that the center of the map will have a traffic jam, the user observes the radar glyphs in Radar Glyph View to see which neighbor areas contribute to the congestions. As shown in \fzz{Fig.~\ref{fig:case}($d$)}, clusters \#1 and \#2 are located in the center of the map.  According to the radar glyphs~(\fzz{Fig.~\ref{fig:case}($c_2$})), the traffic flow will increase 20 minutes later, and the neighbor areas at the north have the most effective correlation. The user can prevent congestion roughly and temporarily by constraint the traffic flow in these areas.

The user takes further steps and handles the latent traffic jam more efficiently and precisely through the visualization and interpretation in Fine-grained Grid View~(\fzz{Fig.~\ref{fig:case}($c_3$,$c_4$})) For example, from the parallel coordinates plot~(\fzz{Fig.~\ref{fig:case}($e$)}), he finds some yellow lines will get a sharp increase 20 minutes later, which can be one of the main locations of traffic jams. He then clicks the corresponding grids in Square Chart~(\fzz{Fig.~\ref{fig:case}($c_3$)}) to project top five significant trajectories. The red trajectories means they all promote traffic jams. The user realizes appropriate reaction should be made on Road \#107 and Road \#110 to prevent traffic jam. Moreover, Bi-directional Column Chart shows that trajectories 20-30 minutes before affect prediction most, which means the user needs to pay more attention on the trajectory in that time slot (highlight in \fzz{Fig.~\ref{fig:case}($c$)}).

As the demo can provide different predictions and corresponding interpretations in three visualization views and easy switching between them, the user is encouraged to customize the analysis workflow himself.

\section{Conclusions and Future Work}

In this work, we present a visual analysis system, TrafPS, to interpret the prediction made by complex models, which consists of three layers: Data processing layer, Prediction-interpretation layer, and Visualization analytical layer. We devise Map-Trajectory View, Radar Glyph View, and Fine-grained Grid View in the Visualization analytical layer to support the analysis and interpretation process interactively. In the demonstration, users can operate TrafPS on real-world taxi data and analyze the effective way to avoid traffic jams. Currently, our work still has some limitation that needs improvement in future work. First, adding more interaction interfaces in the TrafPS to let users adjust and understand interpretation intuitively. Second, implement other division types based on more features with more data. Third, further study is required to get feedback on the visualization system from domain experts and normal users.

\end{document}